\documentclass[prl,twocolumn,showpacs]{revtex4}
\usepackage{graphicx,epsf}

\begin{document}
\title{Ehrenfest time and the coherent backscattering off ballistic cavities}

\author{Saar Rahav}
\affiliation{Laboratory of Atomic and Solid State Physics, Cornell University, Ithaca 14853, USA.}
\author{Piet W. Brouwer}
\affiliation{Laboratory of Atomic and Solid State Physics, Cornell University, Ithaca 14853, USA.}

\date{\today}

\begin{abstract}
If the Ehrenfest time $\tau_{\rm E}$ of a ballistic cavity is not negligible in
comparison to its dwell time $\tau_{\rm D}$, the weak localization
correction to the
cavity's transmission is suppressed proportional to $\exp(-\tau_{\rm
  E}/\tau_{\rm D})$. At the same time, quantum interference enhances
the probability of reflection into the mode of incidence by a factor
two. This `enhanced backscattering' does not depend on the
Ehrenfest time. We show that, in addition to the diagonal enhanced
backscattering, there are off-diagonal contributions to coherent
backscattering that become relevant if $\tau_{\rm E} \gtrsim \tau_{\rm
  D}$.

\end{abstract}
\maketitle

Weak localization and enhanced backscattering are two signatures of
quantum interference in disordered or chaotic conductors
\cite{kn:leshouches1994,kn:beenakker1997}.
Both refer to transmission and reflection
averaged over an ensemble, {\em e.g.}, obtained by slight variations
of the sample shape. Weak localization is a small negative interference
correction to the transmission; Enhanced backscattering is
the phenomenon that the probability of reflection into the incident
scattering mode is twice the probability of
reflection into a different mode. Both weak localization and enhanced 
backscattering require the presence of time-reversal symmetry.
In this letter we consider these phenomena for a two-dimensional 
cavity or `quantum dot' with ballistic and chaotic classical
dynamics. For ballistic cavities the underlying classical dynamics
is known to play an important role determining the quantum
interference effects \cite{kn:stone1995}.


Generally, weak localization and enhanced backscattering are very 
closely related phenomena. There is a particularly simple argument
making this point for the case of a chaotic cavity considered here
\cite{kn:argaman1995}. The chaotic dynamics
is taken to imply that averages of the squares of elements of
the cavity's transmission and reflection matrices 
$t$ and $r$ are equal, except for
the squares of the diagonal elements of $r$, which
are a factor two larger because of enhanced backscattering
\cite{kn:lewenkopf1991,kn:doron1991,kn:baranger1993},
\begin{equation}
  \langle | r_{nn} |^2 \rangle = 2 \langle | r_{mn} |^2 \rangle =
  2 \langle |t_{kn} |^2 \rangle, 
  \ \ \mbox{if $m \neq n$}.
  \label{eq:Snn}
\end{equation}
Here $n$ and $m$, which label modes in the left contact, can take 
the values $n,m=1,2,\ldots,N_1$, where $N_1$ is the number of
propagating modes in 
that contact,
and $k$ takes the values $k=1,\ldots,N_2$, where $N_2$ is the number
of modes in the right contact. 
Since unitarity gives 
\begin{equation}
  \sum_{m=1}^{N_1} |r_{mn}|^2 + \sum_{k=1}^{N_2} |t_{kn}|^2 = 1,
  \label{eq:unitary}
\end{equation}
one finds
that the average of the square of a transmission matrix element is
\begin{equation}
  \langle | t_{kn} |^2 \rangle = \frac{1}{N + 1}, \ \
  N = N_1 + N_2
  \label{eq:Smn}
\end{equation}
The amount by which $\langle |t_{kn} |^2 \rangle$ is smaller than
$1/N$ is the weak localization correction.
\begin{figure}
\epsfxsize=0.95\hsize
\epsffile{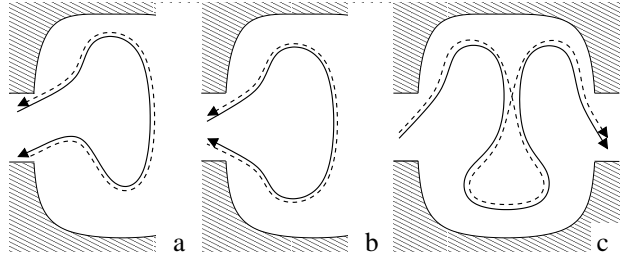}
\caption{\label{fig:1} Schematic drawing of pairs of classical trajectories
  contribution to coherent backscattering (a and b) and weak
  localization (c).}
\vspace{-0.5cm}
\end{figure}

In order to understand the semiclassical origin of enhanced
backscattering, mere inspection
of the semiclassical formula for the square of a reflection matrix
element $|r_{mn}|^2$ is sufficient. Indeed, one has
\cite{kn:jalabert1990}
\begin{equation}
  |r_{mn}|^2 = \frac{1}{2 N_1} \sum_{\alpha,\beta}
  \frac{1}{M_{\alpha} M_{\beta}^*}
  e^{i({\cal S}_{\alpha}-{\cal S}_{\beta})/\hbar},
  \label{eq:Ssemi}
\end{equation}
where $\alpha$ and $\beta$ denote classical
trajectories that connect the left
contact to itself, start with initial
transverse momentum compatible with mode $n$, and end with transverse
momentum compatible with mode $m$, ${\cal S}$ is the classical action
(to be specified later), and $M_{\alpha}$ and $M_{\beta}$ are stability
amplitudes. The modes in each contact have quantized transverse momentum
\begin{equation}
  p_{\perp}(m) = \pm \pi \hbar m/W_{j},\ \ m=1,\ldots,N_{j},
\end{equation}
where $W_j$ is the width of the contact and the subscript $j=1,2$
refers to the left and right contacts, respectively. Enhanced
backscattering then follows from
the simple observation that whereas generically only contributions 
$\alpha = \beta$ contribute to $\langle
|r_{nm}|^2 \rangle$ if $n \neq m$, $\langle |r_{nn}|^2 \rangle$ also 
has contributions from $\alpha$ and $\beta$ being
time-reversed trajectories
\cite{kn:doron1991,kn:lewenkopf1991,kn:baranger1993}. 
Examples of pairs of time-reversed 
trajectories are shown in Fig.\ \ref{fig:1}a and b. 
Since actions and stability amplitudes are
invariant under time reversal, one immediately arrives at the first
equality in Eq.\ (\ref{eq:Snn}).

The semiclassical theory of weak localization is more complicated than
this. As was first shown by Aleiner and Larkin 
\cite{kn:aleiner1996}, weak
localization in ballistic cavities is rooted in destructive
interference of two trajectories that differ only near a small-angle
self-encounter, where one trajectory intersects itself and the other
trajectory does not, see Fig.\ \ref{fig:1}c. 
Since the two trajectories are close everywhere in phase space (up 
to time reversal), their action difference is small.
Aleiner and Larkin's original theory was formulated in a
language borrowed from the theory of disordered conductors; Richter
and Sieber showed that the same structure translates to the
trajectory-based picture of Eq.\ (\ref{eq:Ssemi})
\cite{kn:richter2002}
(see also Ref.\ \onlinecite{kn:heusler2005}).

Aleiner and Larkin's theory of weak localization in ballistic cavities
not only solved a
problem, it also pointed to a new one: Since weak localization in a ballistic
cavity involves the exponential divergence and convergence of
trajectories, the complete theory of weak localization requires
knowledge of the Lyapunov exponent $\lambda$ of the classical dynamics
in the cavity \cite{kn:aleiner1996}. 
The Lyapunov exponent enters the weak localization
correction through the Ehrenfest time $\tau_{\rm E}$, which is the
time it takes for two classical trajectories initially separated by 
a phase space distance $\sim \hbar$ to diverge and be separated by a 
classical phase space distance \cite{kn:zaslavsky1981},
\begin{equation}
  \tau_{\rm E} = \frac{1}{\lambda} \ln N + \mbox{const},
\end{equation}  
where the added term does not scale with Planck's constant. (The
channel number $N$ is proportional to $\hbar^{-1}$.) Since a
small-angle encounter contributing to weak localization has a 
duration $\sim \tau_{\rm E}$, 
the weak localization correction is proportional to 
$\exp(-\tau_{\rm E}/\tau_{\rm D})$, where $\tau_{\rm D}$ is the
cavity's mean dwell time
\cite{kn:aleiner1996,kn:adagideli2003,kn:rahav2005}. Enhanced
backscattering, on the other hand, involves no
small-angle intersections, hence it has no dependence on $\tau_{\rm
  E}$.

Both results, exponential suppression of weak localization with
increasing Ehrenfest time and the Ehrenfest-time independence of
the factor two enhancement of $\langle |r_{nn}|^2 \rangle$ over
generic off-diagonal reflection coefficients $\langle |r_{nm}|^2
\rangle$, have a solid foundation, in
semiclassical theory
\cite{kn:doron1991,kn:lewenkopf1991,kn:aleiner1996,kn:adagideli2003,kn:rahav2005}
as well as numerical simulations \cite{kn:rahav2005,kn:tajic2004}. 
How can they be reconciled? 

The answer to the puzzle, which will be derived below, is that 
coherent
backscattering involves not only the factor-two enhancement of
averaged
diagonal reflection coefficients, but also a slight reduction of 
off-diagonal elements of the reflection matrix if 
$\tau_{\rm E} \gtrsim \tau_{\rm D}$. For the off-diagonal 
elements we find, to leading order in $1/N$,
\begin{eqnarray}
  \label{eq:dr}
  \langle |r_{mn}|^2 \rangle &=&
  \frac{1}{N} 
  \left( 1
  - \frac{1}{2 \lambda \tau_{\rm D}
  |m-n|^{1+1/\lambda\tau_{\rm D}}}\right).
\end{eqnarray}
The reduction of near-diagonal elements in Eq.\ (\ref{eq:dr})
explains why the simple connection between weak localization and
enhanced backscattering used to derive Eq.\ (\ref{eq:Smn}) above
fails if $\tau_{\rm E}$ is not negligible.
One verifies that summation over all off-diagonal reflection
matrix elements gives a contribution $[-1 + \exp(-\tau_{\rm
E}/\tau_{\rm D})]/N$ to the total reflection probability, which
precisely replaces the $\tau_{\rm E}$-independent interference
correction from $\langle |r_{nn}|^2 \rangle$ with the
same $\tau_{\rm E}$-dependence as the weak localization correction
\cite{kn:rahav2005,kn:jacquod2006}.
In Eq.\ (\ref{eq:dr}) we omitted terms that decay proportional
to $|m-n|^{-2}$ or faster and have no net contribution to the total
reflection. The full expression for $\langle |r_{nm}|^2 \rangle$ is
obtained by adding Eqs.\ (\ref{eq:rclass}), (\ref{eq:delta}), and
(\ref{eq:deltaprime}) below.

\begin{figure}
\epsfxsize=0.95\hsize
\epsffile{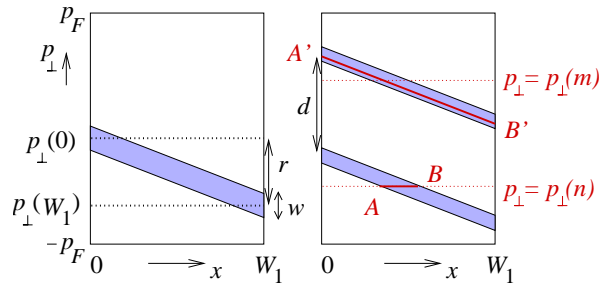}
\caption{\label{genpair} Left: the surface of section of a band,
  together with the definitions of the band's center momentum,
  momentum range $r$, and momentum width $w$. The maximal transverse
  momentum is the Fermi momentum $p_F$. Right: a band and its
  image. The image band is the surface of section of the trajectories
  in the band upon exit, with the transverse momentum $p_{\perp}$
  reversed.}
\vspace{-0.6cm}
\end{figure}

We now describe the details of our calculation. In order to obtain
simple, non system-specific results, we examine the chaotic cavity in
the limit $N \to \infty$ ({\em i.e.}, $\hbar \to 0$), while 
keeping 
$\tau_{\rm E}/\tau_{\rm D}$ fixed. 
In this limit, 
the widths of the contacts scale to zero, $W_j \propto 1/\ln N$,
$j=1,2$. The classical trajectories entering the cavity through the left
contact 
are indexed with the help of their 
transverse momentum $p_{\perp}$ and the coordinate $x$ along the contact's
cross section. 
Alternatively, trajectories exiting the cavity
are indexed with $x$ and {\em minus} their transverse 
momentum. On this `Poincar\'e surface of section', 
trajectories appear in `bands' that remain close in phase space 
throughout the passage through the cavity 
\cite{kn:wirtz1999,kn:tworzydlo2003}. The phase space points 
that belong to a
band are concentrated along stable manifolds at entry and unstable
manifolds at exit. The band edges are determined by the contact edges.
Since the widths $W_1$ and $W_2$ scale to zero in the classical
limit taken here, we can linearize the 
dynamics within each band. While trajectories inside a
band have correlated actions, we 
assume that actions of trajectories in different bands are
uncorrelated. (This precludes a theory of weak
localization.) Linearization of the classical dynamics inside a
band also means that the edges of nearby bands are parallel
lines in the Poincar\'e surface of section. For simplicity, we 
assume that all relevant bands are parallel; we neglect `folds', bands with a
strong curvature that do not span the entire range $0 < x < W_1$.

For each band in the Poincar\'e surface of section referencing the
trajectories at their entrance into the cavity, there exists an `image
band' in the Poincar\'e surface of section referencing the
trajectories at their exit. 
This `entrance-exit' relation maps the Poincar\'e 
surface of section onto itself,
so that one obtains a connection between pairs of time-reversed
bands. (Although bands may be mapped onto themselves, only pairs
of different bands can give rise to interference.)

Each band is characterized by its center momentum $p(0)$ at
$x=0$, its center momentum $p(W_1)$ at $x=W_1$ and its momentum width
$w$, see Fig.\ \ref{genpair}.
(Note that it is implicitly assumed that the cross section is at the lead opening.)
For a band, we define its momentum range as $r = |p(0) -
p(W_1)|$. For a pair of bands,
we define their distance $d$ as the difference of the center momenta. 
Both bands in a pair have the same width \cite{kn:tworzydlo2003}.
Since trajectories originating from bands a distance $d$ apart are
correlated for a time $\lambda^{-1} \ln p_F/d$
\cite{kn:aleiner1996,kn:zaslavsky1981}, 
the dwell time $t$ for trajectories in a band a
distance $d$ from its image is at least $t_{\rm min}$,
with 
\begin{equation}
  t_{\rm min} = (2/\lambda) \ln (p_F/d), \label{eq:dbound}
\end{equation}
where $p_F$ is the Fermi momentum. As the momentum width $w$ of a band 
scales $\propto \exp(-\lambda t)$ \cite{kn:wirtz1999}, 
we have $w \ll d,\ r$ for a typical pair of bands and for dwell times 
of order $\tau_{\rm D}$ and larger. This allows us to neglect
the momentum width of bands $w$ with respect their distance $d$ or
momentum range $r$ in the following considerations.

It is important to note that in each band there is at most one
trajectory with given transverse momenta for entrance and exit. This
is illustrated in the right panel of
Fig.\ \ref{genpair}. A trajectory which starts with
transverse momentum $p_{\perp} = p_{\perp}(n)$ 
starts somewhere on the linear segment AB in the
figure, which is the intersection of the band and the line 
$p_{\perp} = p_{\perp}(n)$.
Since the phase space points A and B correspond to the band's
edge, they are mapped to the edges $x=0$ and $x=W_1$ of the left
contact upon exit: The image of the 
segment AB is the line A'B' in the right panel of Fig.\ \ref{genpair}.
For
an (inverted) outgoing transverse momentum $p_{\perp}(m)$ there is 
at most one intersection with A'B'. Note that there is precisely one 
solution if $p_{\perp}(m)$ is in the momentum range of the image 
band and if we may ignore the possibility that the lines $p_{\perp} =
p_{\perp}(n)$ or $p_{\perp}=p_{\perp}(m)$ cut the bands close to $x=0$
or $x=W_1$. This simplification is allowed since in the limit
considered here all bands are narrow, see the discussion following Eq.\ 
(\ref{eq:dbound}).

Having found a way to index families of trajectories appearing in the
semiclassical formula (\ref{eq:Ssemi}), we now describe the
calculation of the contribution of such trajectories to $\langle
|r_{mn}|^2 \rangle$. 
Hereto, we need to specify the action ${\cal S}$ entering into Eq.\
(\ref{eq:Ssemi}) \cite{kn:jalabert1990},
\begin{equation}
\label{mods}
 {\cal S} = \tilde{\cal S} + p_{\perp} x -
 p^\prime_{\perp} x^\prime
\end{equation}
where $\tilde{\cal S}$ is the standard coordinate-dependent 
classical action \cite{kn:jalabert1990} and unprimed and 
primed variables refer to entrance and exit, respectively. The last
two terms in Eq.\ (\ref{mods}) are obtained from the lead wave functions
and cause ${\cal S}$ to depend explicitly on the precise location
of the contacts. 
In the linearized formulation used here, the stability
amplitude $M$ is the same for all trajectories in a band. Neglecting the
band width $w$ with respect to its momentum range $r$, $|M|^2$ 
can be estimated as
\begin{equation}
  |M|^2 = r^2/p_F w.
  \label{eq:malpha}
\end{equation}

Before calculating the quantum-interference corrections from pairs
of time-reversed bands, we calculate the leading classical
contribution to $\langle |r_{mn}|^2 \rangle$ using the band 
picture. The classical contribution to $\langle |r_{mn}|^2 \rangle$
follows from taking $\alpha = \beta$ in Eq.\ (\ref{eq:Ssemi}).
In order to find $\langle |r_{mn}|^2 \rangle$ we need to sum over 
over all reflection bands for which the initial momentum range contains
$p_{\perp} = \pi \hbar n/W_1$ or $p_{\perp} = - \pi \hbar
n/W_1$ and for which the image band contains $p_{\perp} = 
\pi \hbar m/W_1$ or $-\pi \hbar m/W_1$. The number density 
$n(t)$ of bands with this property and with dwell time $t$ 
is obtained by multiplying the probability density 
$(N_1/N \tau_{\rm D}) \exp(-t/\tau_{\rm D})$ for escape at
time $t$ through contact $1$ by the probability $(r/p_F)^2$ that
the band and the image band intersect one of the lines $p_{\perp} = 
\pm \pi \hbar n/W_1$ and $p_{\perp} = \pm \pi \hbar m/W_1$,
respectively, and dividing by the area fraction of one band,
$w/2 p_F$, 
\begin{equation}
  n(t) = \frac{2 N_1 r^2}{N w p_F \tau_{\rm D}} e^{-t/\tau_{\rm
  D}}.
  \label{eq:nt}
\end{equation}
Dividing $n(t)$ by $2 N_1 |M|^2$
and integrating over time, we find the well-known leading-order result
\begin{equation}
  \langle |r_{mn}|^2 \rangle = 1/N.
  \label{eq:rclass}
\end{equation}

There are two distinct types of coherent backscattering corrections to
Eq.\ (\ref{eq:rclass}): From trajectories which
start and end with approximately opposite transverse momenta (as in 
Fig.\ \ref{fig:1}a) and from
trajectories for which incoming and outgoing transverse
momenta are approximately
equal (as in Fig.\ \ref{fig:1}b). We address the former 
contribution first.

In order to have an interference contribution from trajectories with
approximately opposite incoming and outgoing momenta, the momentum
distance $d$ of their band and image band must be less 
than the range $r$ of each band:
only then there exist values of the incoming and outgoing momenta
that intersect both bands.
(Since we take bands to be parallel, 
both have the same momentum range $r$.)
For each pair of
bands, there are two trajectories with incoming transverse momentum
$\pm \pi n \hbar/W_1$ and outgoing transverse momentum $\mp \pi m \hbar/W_1$. 
If $n = m$
these two trajectories are time reversed and have the same action ${\cal
  S}$. Their interference gives half the standard enhancement of
$\langle |r_{nn}|^2 \rangle$. (The other half comes from the
class of trajectories considered next.) If $n \neq m$ the 
action difference 
is calculated to be
\begin{equation}
  \Delta {\cal S} = \pi \hbar (m-n) r^{-1} d. \label{eq:dS}
\end{equation}
In calculating Eq.\ (\ref{eq:dS}) we used the linearity of the band
and neglected the width of a band, $w$, in comparison to 
$d$ and $r$. 
The sign of the action difference is
irrelevant, since the sum over trajectories (\ref{eq:Ssemi})
contains both $\Delta {\cal S}$ and $-\Delta {\cal S}$. In order to
find the corresponding interference contribution $\delta_{|m-n|}$
to $\langle |r_{nm}|^2 \rangle$, we need the number density $n(t,d)$ 
of reflection bands that have distance $d$ and
intersect both $p_{\perp} = \pm n \pi \hbar/W_1$ and $p_{\perp} = \pm
m \pi \hbar/W_1$,
$$
  n(t,d) = 
  \frac{N_1 (r - |m-n|\hbar \pi/W_1 - d)}{N w p_F \tau_{\rm D}}
  e^{-(t-t_{\rm min}/2)/\tau_{\rm D}},
$$
if $0 < d < r - |m-n|\hbar \pi/W_1$ and $t > t_{\rm min}(d)$; $n(t,d)
= 0$ otherwise.
Both the minimum dwell time $t_{\rm min}$ and the exponential
enhancement factor $\exp(t_{\rm min}/2)$ follow from the correlations of
trajectories in nearby bands, see Eq.\ (\ref{eq:dbound}) and Refs.\
\onlinecite{kn:heusler2005} and \onlinecite{kn:rahav2005}. 
Integrating over the dwell time $t$ and the band distance $d$, we then
find
\begin{equation}
  \delta_l =   \int_0^{l-y(l)} dz
  \frac{l-y(l)-z}{l^{2}N}
  (z/l)^{1/\lambda \tau_{\rm D}} \cos(\pi z),
  \label{eq:delta}
\end{equation}
where $y(l) = l^2 \pi \hbar/W_1 r$.
  
Contributions of trajectories for which the incoming and
outgoing transverse momenta are approximately equal
are treated similarly. The main difference with the preceding
contribution is that, in this case, small action differences are found
for pairs of bands that are typically far from each other.
The action difference can be estimated in exactly the same way as it 
was done previously,
\begin{equation}
  \Delta {\cal S}= \pi \hbar (m-n) (p+p') r^{-1},
\end{equation}
where $p$ and $p'$ are the center momenta of the two bands in the
pair taken at $x=0$ or $x=W$. We then find that the interference 
correction to $\langle |r_{nm}|^2 \rangle$ from this class of
trajectories is $\delta'_{|m-n|}$, with \cite{foot2}
\begin{equation}
  \delta'_l = 
  \int_{y(l)}^{l} dz
  \frac{z - y(l)}{l^2 N} \cos(\pi z).
  \label{eq:deltaprime}
\end{equation}

Equations (\ref{eq:delta}) and (\ref{eq:deltaprime}) contain the main
quantitative result of this letter. Summation over $l$ gives the 
coherent backscattering correction 
to the total reflection \cite{kn:rahav2005}, which now depends on
the Ehrenfest time in the same way as the weak localization correction
and, together with weak localization, preserves the unitarity relation 
(\ref{eq:unitary}).
Combining both interference contributions and evaluating the
integrals in the limit $\hbar \to 0$ and for large
$\lambda \tau_{\rm D}$ one finds Eq.\
(\ref{eq:dr}) above with the extra condition that $|m-n|$ be small 
in comparison to $r W_1/\hbar$, plus a contribution that decays faster with $l$
than Eq.\ (\ref{eq:dr}) and sums to zero if summed over $l$ in the
limit $\hbar \to 0$ at fixed $\tau_{\rm E}/\tau_{\rm D}$.
While Eqs.\ (\ref{eq:delta}) and (\ref{eq:deltaprime}) were derived
under the assumption that all reflection bands are parallel in the 
Poincar\'e surface of section, we verified that relaxing this
assumption for Eq.\ (\ref{eq:deltaprime}), where it is most
questionable, only gives corrections to $\delta'_l$ that are of
order $1/N^2$ and sum to zero in the limit
$\hbar \to 0$ at fixed $\tau_{\rm E}/\tau_{\rm D}$.

In conclusion, we have shown that there are off-diagonal contributions 
to coherent backscattering off a ballistic chaotic cavity
which are essential if the Ehrenfest time
$\tau_{\rm E}$ is not small in comparison to the dwell time $\tau_{\rm D}$.
To the best of our knowledge, the Ehrenfest-time dependence of the
average reflection coefficients $\langle |r_{nm}|^2 \rangle$
is the first manifestation of the
Ehrenfest-time in a quantity that does not involve a summation over
lead modes. As such, coherent backscattering not only probes
the Ehrenfest regime for ballistic electronic transport, it also 
provides a method to access the Ehrenfest regime for experiments 
with classical waves, such as microwave billiards.

We thank Carlo Beenakker, Douglas Stone, and Denis Ullmo for stimulating
discussions. The unpublished Ref.\ \onlinecite{kn:tajic2004} motivated
us to start this
research. Upon completion of this work Refs.\
\onlinecite{kn:jacquod2006} and \onlinecite{kn:whitney2006}
appeared on the cond-mat archive, in which
the off-diagonal nature of coherent backscattering was also pointed 
out. We thank Philippe Jacquod for sending us a preprint of Ref.\
\onlinecite{kn:jacquod2006}. 
This work was supported by the NSF under grant no.\ DMR 0334499 and 
by the Packard Foundation.

\vspace{-0.5cm}


\begin{thebibliography}{16}
\bibitem{foot2}
The main contribution of Eq.\
  (\ref{eq:deltaprime}) to the total reflection, though
negligible in the classical limit taken here, is from pairs of
bands with $p+p'\approx 0$ or $p+p' \approx \pm 2 r$. This describes the effect of
diffraction at the contact edges.
\end{thebibliography}

\end{document}